\documentclass[submission,copyright,creativecommons]{eptcs}
\providecommand{\event}{ACL2 2025} 

\usepackage{iftex}

\ifpdf
  \usepackage{underscore}         
  \usepackage[T1]{fontenc}        
\else
  \usepackage{breakurl}           
\fi

\usepackage[dvipsnames]{xcolor}

\usepackage{listings}
\usepackage[scaled=0.85]{beramono}            
\definecolor{shadecolor}{rgb}{.9, .9, .9}

\lstdefinestyle{plain}
{ 
  columns=fixed,
  frame=none,
   xleftmargin=.15\textwidth, xrightmargin=.15\textwidth
}

\lstdefinestyle{acl2}
{basicstyle=\footnotesize\ttfamily}

\usepackage{graphicx}
\graphicspath{{./graphics}}

\title{An ACL2s Interface to Z3}
\author{Andrew T. Walter \qquad\qquad Panagiotis Manolios
  \institute{Khoury College\\
    Northeastern University\\
  Massachusetts, USA}
\email{walter.a@northeastern.edu \quad\qquad p.manolios@northeastern.edu}
}
\def\titlerunning{An ACL2s Interface to Z3}
\def\authorrunning{A.T. Walter \& P. Manolios}
\begin{document}

\newcommand{\lispsmt}{Lisp-Z3}
\newcommand{\ea}[0]{\emph{et al.}}
\newcommand{\acls}[0]{ACL2/s}
\newcommand{\p}[1]{\texttt{#1}}

\newcommand{\ie}{\emph{i.e.}}
\newcommand{\eg}{\emph{e.g.}}

\maketitle

\begin{abstract}
  We present \lispsmt, an extension to the ACL2s systems programming
  framework (ASPF) that supports the use of the Z3 satisfiability
  modulo theories (SMT) solver. \lispsmt\ allows one to develop tools
  written using the full feature set of Common Lisp that can use both
  ACL2/s (either ACL2 or ACL2s) and Z3 as services, combining the
  power of SMT and interactive theorem proving. \lispsmt\ is usable by
  anyone who would like to interact with Z3 from Common Lisp, as it
  does not depend on the availability of ACL2/s. We discuss the use of
  \lispsmt\ in three applications. The first is a Sudoku solver.
  The second is SeqSolve, a string solver which solved a larger number
  of benchmark problems more quickly than any other existing solver at
  the time of its publishing. Finally, \lispsmt\ was also used in the
  context of hardware-in-the-loop fuzzing of wireless routers, where
  low latency was an important goal. The latter two applications
  leveraged the ability of \lispsmt\ to integrate Z3 with ACL2s
  code. We have further plans to use \lispsmt\ inside of ACL2s to
  provide more powerful automated support for dependent types, and in
  particular more efficient generation of counterexamples to
  properties involving dependent types.
  This paper describes the usage and implementation of \lispsmt, as
  well as an evaluation of its use in the aforementioned applications.
\end{abstract}

\section{Introduction}
\label{sec:intro}

This paper describes a publicly available extension to our ACL2s
systems programming framework~\cite{acl2s-systems-programming} (ASPF)
that supports the use of the Z3 satisfiability modulo theories (SMT)
solver~\cite{z3} as a service.

ASPF enables the development of tools that use ACL2 and ACL2s (the
ACL2 Sedan) as a service by allowing one to write code that uses
Common Lisp features that \acls\ (ACL2 and ACL2s) restrict. This code
can then interact with \acls\ using a library provided by ASPF. We
have used ASPF to build several systems, including a web-based loop
invariant discovery game~\cite{invariant-discovery-game}, a system for
providing feedback for calculational proofs intended for pedagogical
settings~\cite{acl2-workshop-checker-paper} and a system for
automating the grading of homework involving different kinds of
automata~\cite{automata-grading}. In our experience, ASPF particularly
shines when building tools that are components of a larger system,
especially when networking and foreign-function interfacing (FFI) are
required.


Z3 is an SMT solver. This means that given a set of constraints within
a supported theory, Z3 will attempt to determine whether or not that
set of constraints is satisfiable. If so, Z3 can produce a satisfying
assignment (a \emph{model}) for the constraints. Z3 may be able to
determine that the constraints are unsatisfiable as well, or it may
instead exceed a timeout or resource limit and report that the
satisfiability of the constraints is unknown. \lispsmt\ provides an
interface for expressing and asserting constraints, requesting that Z3
check the satisfiability of asserted constraints and accessing the
produced satisfying assignment if Z3 determined that the constraints
were satisfiable. The interface of \lispsmt\ is intended to mirror the
SMT-LIB2~\cite{smtlib} command interface as much as possible, making
it especially easy to use for anyone who has experience with SMT-LIB2
(which Z3 and many other SMT solvers support). Many kinds of problems
can be modeled using SMT, with a classic example being solving Sudoku
puzzles.


As prior work has
reported~\cite{framework-verifying-pipelined-machines, smtlink,
  smtlink-two}, interactive theorem proving (ITP) and SMT are
complementary techniques and their combination can be highly
effective. The authors and their collaborators have found the
combination of \acls\ (which is an interactive theorem prover) and Z3 to
be useful in multiple applications, including string solving (Kumar
\ea's TranSeq)~\cite{ankit-mpmt} and security testing of wireless
routers~\cite{enumerative-data-types}. To support these applications,
it was necessary to develop an \acls\ or Common Lisp interface for Z3
with the right features---in the case of security testing, low latency
was highly desirable, whereas incremental solving was important for
the string solver. Existing interfaces did not fulfill these
requirements. Our interface, which we call \lispsmt, consists of
low-level bindings to Z3's C API as well as a higher-level interface
on top to make it convenient to interact with Z3. \lispsmt\ is usable
by anyone who would like to interact with Z3 from Common Lisp, as it
does not rely on functionality specific to \acls. Nevertheless, we
think of \lispsmt\ as an extension of the ACL2s Systems Programming
methodology, providing another reasoning backend in addition to \acls.


In addition to using \lispsmt\ when developing tools that use \acls\
as a service, we are planning to use \lispsmt\ to power functionality
inside of ACL2s. In particular, we are working on Enumerative Data
Types Modulo Theories, a generalization and extension of our wireless
router security testing project~\cite{enumerative-data-types} that
aims to improve the ability of ACL2s to generate counterexamples in
the presence of
constraints. 
This will involve using \lispsmt\ inside of ACL2s' cgen~\cite{cgen},
which is integrated into the ACL2 waterfall.

The contributions of this work include: \textbf{(1)} A description of
the design and implementation of \lispsmt, a major extension to ASPF
that in addition to supporting ACL2s also supports Z3. With this
extension, one can build tools that use both ACL2/s and Z3 as
services, \textbf{(2)} A public release of the extended ASPF,
including examples of the use of \lispsmt\ in Common Lisp outside of
ACL2/s, \textbf{(3)} an evaluation of the use of \lispsmt\ in
conjunction with the ASPF in three applications: a Sudoku solver, a
state-of-the-art string solver and hardware-in-the-loop fuzzing of
wireless routers.

The remainder of the paper is organized as follows:
Section~\ref{sec:usage} gives a brief introduction to the interface of
\lispsmt\ through examples, Section~\ref{sec:intro-z3-smtlib} gives
short introduction to Z3 and SMT-LIB2,
Section~\ref{sec:implementation} discusses how \lispsmt\ is
implemented, Section~\ref{sec:sudoku} walks through the development of
a Sudoku solver using \lispsmt, Section~\ref{sec:string-solving}
discusses the use of \lispsmt\ in the SeqSolve string solver,
Section~\ref{sec:wifi} explores the use of \lispsmt\ in
hardware-in-the-loop fuzzing of wireless routers,
Section~\ref{sec:related-work} provides an overview of related work,
and Section~\ref{sec:conclusion} concludes.

\section{Usage}
\label{sec:usage}

Listing~\ref{lst:basic-usage-smtlib} shows a basic example of the
usage of \lispsmt. After initializing Z3, the variables \lstinline|x|
and \lstinline|y| are declared in the same way that one might declare
them in SMT-LIB2, using the \lstinline|declare-const|
command. \lstinline|x| is declared to be a Boolean variable, and
\lstinline|y| is declared to be an integer variable. Next, the
\lstinline|z3-assert| function is used to add a constraint to Z3. The
constraint added states that \lstinline|x| must be true, and that
\lstinline|y| must be greater than or equal to 5. Note that this could
be written as two independent calls to \lstinline|z3-assert| rather
than as a conjunction if desired. Next, the \lstinline|check-sat|
command is run, which asks Z3 to determine whether the conjunction of
all of the constraints added to it is satisfiable. Here it will return
\lstinline|:SAT|, indicating that the constraints are indeed
satisfiable. Finally, we call \lstinline|get-model| to retrieve Z3's
representation of a satisfying assignment to the free variables in the
constraint we added. In this case, Z3's representation of one possible
satisfying assignment is printed as follows:

\begin{lstlisting}[style=acl2, label=lst:basic-example-assignment, floatplacement=h]
#<Z3::MODEL 
  X -> true
  Y -> 5    >
\end{lstlisting}

If one would like to interact with the assignment in Common Lisp, it
is generally easier to instead call
\lstinline|get-model-as-assignment|, which will translate Z3's
representation of the satisfying assignment into a Common Lisp list
appropriate for use as let bindings. In this case, the output would be
\lstinline|((X T) (Y 5))|. Note that there are infinitely many
satisfying assignments to this set of constraints, as \lstinline|y|
may be assigned any integer greater than 5. In principle Z3 could
produce any of these assignments, though in this case it tends to
generate the solution shown above, which is the satisfying assignment
with the smallest possible value for \lstinline|y|.

\begin{lstlisting}[style=acl2, label=lst:basic-usage-smtlib, caption={An example of a basic SMT query using \lispsmt, in the style of SMT-LIB2.}]
;; Load lisp-z3
(ql:quickload :lisp-z3)
;; Enter its package so we can use its functions without needing to
;; specify the z3 package.
(in-package :z3)
;; Set up Z3. Only needs to happen once, before other code that uses Z3
(solver-init)
;; Declare variables x and y
(declare-const x Bool)
(declare-const y Int)
;; Add an assertion
(z3-assert
  (and x (>= y 5)))
;; Check for satisfiability
(check-sat)
;; If satisfiable, get a satisfying assignment
(get-model)
\end{lstlisting}

\lispsmt\ also allows one to declare variables inline with the
\lstinline|z3-assert| form. This is shown in
Listing~\ref{lst:basic-usage-inline}. This syntax is similar to that
used by ACL2s' \lstinline|definec| and \lstinline|property| forms,
making it easier for users familiar with those facilities to start
using \lispsmt.

\begin{lstlisting}[style=acl2, label=lst:basic-usage-inline, caption={An example of a basic SMT query using \lispsmt, using inline declarations of variables rather than forward declarations as shown in Listing~\ref{lst:basic-usage-smtlib}.}]
;; Set up Z3. Only needs to happen once, before other code that uses Z3
(solver-init)
;; Declare variables x and y and add an assertion over them
(z3-assert (x :bool y :int)
  (and x (>= y 5)))
;; Check for satisfiability
(check-sat)
;; If satisfiable, get a satisfying assignment and translate it into a
;; form that is usable as Common Lisp let bindings
(get-model-as-assignment)
\end{lstlisting}

It is important to note that the statement passed in to
\lstinline|z3-assert| to be asserted in Z3 will be interpreted using
the semantics that Z3 assigns to the used operators. Z3's semantics
for expressions diverge from the semantics of ACL2 in some cases, as
will be discussed later.

\section{Short Introduction to Z3 and SMT-LIB2}
\label{sec:intro-z3-smtlib}

Z3 supports several input formats, but the default is
SMT-LIB2~\cite{smtlib}. SMT-LIB2 was developed with the intention of
creating a standard format for interacting with different SMT
solvers. SMT-LIB2 consists of several components, including a command
language for use when interacting with a SMT solver. All of the
languages that SMT-LIB2 provides are based on S-expressions. The base
logic used in SMT-LIB2 is derived from many-sorted first-order logic
with equality, meaning that functions, variables and operators have
sorts associated with them. In this context, a sort can be thought of
as a name for a type. SMT-LIB2 also provides a standard set of
\emph{theories}, each of which include declarations for the sorts and
functions that the theory provides. For example, the \lstinline|Ints|
theory provides the \lstinline|Int| sort and a set of functions over
\lstinline|Int|s (addition, multiplication, negation, subtraction,
division, modulus, absolute value, and inequality relations).

To express a set of assertions and check its satisfiability using the
SMT-LIB2 command format, one will generally do the following: 1)
declare or define any sorts, functions and constants (variables) that
will be used beyond what is provided by the theory in use; 2)
manipulate the set of assertions maintained by the SMT solver, for
example by adding assertions over the declared sorts, functions and
constants; 3) request that the SMT solver perform a satisfiability
check and print a model. The produced model may not have an
interpretation (an assigned value) for every declared sort, function
and constant from the assertion stack. This generally will occur if
the satisfiability of the assertion stack is not dependent on that
sort, function or constant having a particular value.

SMT-LIB2 solvers maintain an \emph{assertion stack} that consists of
\emph{assertion levels}. Each assertion level is a set containing
assertions as well as declarations of sorts, functions and
constants. When the solver is asked to check satisfiability, it
considers the contents of all of the assertion levels in the
stack. SMT-LIB2 provides commands for manipulating the stack.
\lstinline|push| allows one to create a new assertion level, and
\lstinline|pop| removes the most recently introduced assertion level
from the stack. This removes any of the assertions added since the
popped assertion level was introduced. The behavior of popping on sort
and variable declarations is controlled by the
(\lstinline|:global-declarations|) solver option. If this option is
set to \lstinline|false| (as is the default in Z3), a declaration of a
sort or a variable is attached to the assertion level of the solver at
the time of declaration. If that assertion level is popped off the
stack, the declaration is removed. If the option is set to
\lstinline|true|, declarations of variables and sorts are unaffected
by changes to assertion levels. Maintaining an assertion stack means
that an SMT-LIB2 solver can support a kind of \emph{incremental
  solving}, where satisfiability is queried multiple times, with
modifications made to the set of assertions in between queries.




\section{Implementation}
\label{sec:implementation}

\lispsmt\ consists of two main parts: the low-level bindings to Z3's C
API, and the higher-level interface that provides a convenient
interface for asserting constraints and generally interacting with
Z3. These two parts together make up an ASDF~\cite{asdf-project}
system that can be loaded by many Common Lisp implementations.

\subsection{The Low-Level Interface}

Included in Z3's distribution is a library that can be used to
integrate Z3 inside another program. Z3 provides APIs that allow one
to call into this library from several different programming
languages. We chose to write bindings for the C API provided by Z3, as
C foreign function interfacing (FFI) is common and there is
substantial support available for doing so in Common Lisp. We used the
Common Foreign Function Interface (CFFI) library~\cite{cffi} to
implement our bindings in a way that is portable across many Common
Lisp implementations.

Interfacing with C in Common Lisp results in certain challenges.  For
example, to be able to call a C function that takes in an argument of
type \lstinline|Z3_context|, the Common Lisp implementation needs to
know the size of values of that type, the layout of any fields (if it
is a C struct) and how to turn a Common Lisp value into a
\lstinline|Z3_context| value. Even just determining the size of the
type is a complicated affair, as it generally requires looking at the
C header files where the type is defined, which involves handling
preprocessor directives which may appear in those header files, and
then making a guess as to what size a C compiler would use for values
satisfying that definition. In practice, FFI tools often manage these
issues by generating a C file that includes the relevant types and
interfacing with a C compiler to determine whatever information is
needed about those types. For \lispsmt, we use CFFI's Groveller
functionality. We provide a special Common Lisp file called a
\emph{Grovel file} that has a form for each Z3 type we would like to
interact with. The Groveller evaluates this file to produce a C file
which is then compiled and run. The result of running the resulting
executable is another Lisp file that contains CFFI forms that describe
the layout and size of the Z3 types we referenced. We can then load
the Z3 library and use CFFI forms to create Lisp bindings for the Z3
functions that we would like to call, using the size and layout
information that was gleaned previously.

The Grovel file must be aligned with the API provided by the version
of Z3 running on the user's computer. For example, different versions
of Z3 may provide different members for an enumeration type used to
identify which built-in operator a function call is using. To make it
easier for a user to generate an appropriate version of the Grovel
file, we provide a Python script that will read Z3's C header files
and generate a Grovel file appropriate for them. A similar issue
exists for the file that contains bindings for each Z3 C function that
we would like to expose, though we do not yet provide an automated way
to generate that file. We try to ensure that \lispsmt\ is shipped with
files that should work with a relatively modern version of Z3.
This is done by using the Grovel file generation script and manually
removing or modifying functionality for maximal compatibility.

At this point, it is possible to call many of Z3's C API functions,
but it is not convenient to do so. One needs to manually deal with
memory management tasks, array types are a pain to deal with, printing
values of Z3 types gives little useful information and the context
value must be provided in practically every function call. An example
highlighting the verbosity of the low-level interface is provided in
Listing~\ref{lst:low-level-example}. Note that this example does not
include any error handling and also avoids functionality that requires
manual reference counting (memory management).  This is where the
high-level interface comes in!

\begin{lstlisting}[style=acl2, label=lst:low-level-example, caption={An example highlighting the usage of the low-level interface.}]
;; The below form asserts the constraint (= (+ x y) 10) for integer variables
;; x and y, checks satisfiability and reports a satisfying assignment if SAT.
(let* ((ctx (z3-mk-context (z3-mk-config)))
       (slv (z3-mk-simple-solver ctx))
       (x (z3-mk-const ctx (z3-mk-string-symbol ctx "X") (z3-mk-int-sort ctx)))
       (y (z3-mk-const ctx (z3-mk-string-symbol ctx "Y") (z3-mk-int-sort ctx)))
       ;; add has arbitrary arity, so we need to provide the args in a temporary C array.
       (sum (with-foreign-array (arg-array z3-c-types::Z3_ast (list x y))
              (z3-mk-add ctx 2 arg-array)))
       (stmt (z3-mk-eq ctx sum (z3-mk-numeral ctx "10" (z3-mk-int-sort ctx)))))
  (z3-solver-assert ctx slv stmt)
  ;; Check whether the assertion is satisfiable
  (if (equal (z3-solver-check ctx slv) :L_TRUE)
      ;; SAT! Now we must get all of the constant interpretations (e.g. variable
      ;; assignments) from the model.
      (let ((model (z3-solver-get-model ctx slv)))
        (loop for i below (z3-model-get-num-consts ctx model)
              for decl = (z3-model-get-const-decl ctx model i)
              for name = (z3-get-symbol-string ctx (z3-get-decl-name ctx decl))
              for value-ast = (z3-model-get-const-interp ctx model decl)
              ;; Here we assume the value is a numeral and get it as a string
              collect (list name (z3-get-numeral-string ctx value-ast))))
       ;; Otherwise, UNSAT or unknown.
      'not-sat))
;; Outputs (("Y" "0") ("X" "10"))
\end{lstlisting}

\subsection{The High-Level Interface}

The high-level interface mitigates several of the pain points that the
low-level interface gives rise to. It is written entirely in Common
Lisp and uses the low-level interface internally to make calls to
Z3.

\paragraph*{The Context and Solver}

When interacting with Z3 programatically, one is nearly always doing
so with respect to a particular \emph{context} value. The context
stores certain settings and global values as well as information
needed for memory management (discussed later). Since most operations
on Z3 types require the context that the value was created relative
to, we define a wrapper type around each Z3 type that has a field for
the relevant context in addition to the value itself. In addition to
making it unnecessary for the user to pass a context value around when
dealing with \lispsmt\ code, this makes it possible to implement
\lstinline|describe-object| and \lstinline|print-object| for each Z3
type, enabling Common Lisp to display useful printed representations
for values of Z3 types.

Another important element of the Z3 C API is the \emph{solver} value,
which stores any constraints that the user adds and is needed when
checking satisfiability or generating a satisfying assignment to the
set of constraints. When \lispsmt\ is initialized, a default solver is
created and stored. This solver is used whenever the user does not
specify one. Many parameters of the solver can be modified to control
Z3's behavior---for example, one can set how many threads will be used
by Z3 when checking satisfiability, the logic used to set up the SMT
solver and the schedule used for performing restarts. 


\paragraph*{Memory Management}

As is often the case when interfacing with a C API from a language
with automatic memory management, one must be careful to ensure that
any allocated memory that passes over the language barrier is
deallocated at an appropriate time. For many of the types that it
defines, Z3's C API provides a manual reference counting interface for
managing the lifetime of allocated memory. This means that each time
we create an object that requires manual reference counting, like a
solver, we must call a function to increment the reference counter for
that object. This is implemented for each Z3 type by incrementing the
reference counter in the initializer of the corresponding wrapper
type.  As long as an object's reference counter has a positive value,
Z3 will not deallocate that object's memory. We use the
trivial-garbage Common Lisp library~\cite{trivial-garbage} to attach a
finalizer function to each such object that will run when the wrapper
object has been garbage collected (\eg\ when it is no longer
referenced by any Lisp values). This finalizer function decrements the
object's reference counter so that Z3 is notified that one fewer
reference to the object exists. When the reference counter hits zero,
Z3 can deallocate that object's memory.

\paragraph*{Producing Expressions}

\lispsmt\ aims to support expressing as many of the constraints that
Z3 supports as possible. To assert constraints in a Z3 solver, we
first need to convert them into Z3 AST objects. This may seem trivial,
since the default input format for the Z3 binary is based on
S-expressions, but in practice it is more complicated than simply
handing off an S-expression to Z3.

The primary mechanism that \lispsmt\ provides for expressing
constraints to be asserted in Z3 is the \lstinline|z3-assert|
macro. In addition to taking in an expression to be asserted, this
macro can optionally take in a set of specifiers for free variables to
be used in the assertion, as well as a solver object. Each specifier
contains a name and a sort specification describing the signature of
the variable. These will be described in more detail later. All
assertions are performed with respect to a solver object. If the
solver object is not provided explicitly, the default solver is used.
The assertion is traversed recursively, with each argument of a
function call or operator application being translated into a Z3 AST
before the function call or operator application itself is
translated. Whenever a reference to a free variable is found, an
appropriate Z3 AST object referencing a free variable with the correct
name and sort is created. Information about the set of known
identifiers and their sorts is maintained by the solver. \lispsmt\ has
support for a subset of the operators supported by Z3. The operators
can be referenced by the same name that they are known by in Z3's
SMT-LIB2 interface, though some are known by additional names as well
(\emph{aliases}). A document describing the set of operators known by
\lispsmt\ is provided alongside its source code~\cite{repo}.

The way that SMT-LIB2 behaves in situations where there are multiple
declarations of variables with the same name is different from the way
that Common Lisp does. In particular, SMT-LIB2 provides a single
namespace for variables (constants and functions) and allows multiple
declarations of variables with the same name, given that they are
associated with different sorts. To reference such a variable, it is
necessary to disambiguate using the \lstinline|as| form. For example,
if both a constant of type \lstinline|Int| and a function of type
\lstinline|(Int) -> String| have been declared with the name
\lstinline|x|, one must reference the constant using the form
\lstinline|(as x Int)|. An exception to this behavior is when
variables are introduced by a form that introduces bound variables,
like \lstinline|forall| or \lstinline|exists|. If such a form
introduces a variable with name \lstinline|x|, any references to
\lstinline|x| in the body of that form (unless inside another form
that introduces \lstinline|x| as a bound variable) will refer to the
bound variable rather than any declaration outside of binding form.

To behave in a way that is more consistent with Common Lisp, \lispsmt\
restricts the declarations of variables. In particular, \lispsmt\
requires that at all times, any name is associated with at most one
declared free variable. Declarations of variables are associated with
the solver's assertion level at the time of the declaration and are
removed when that assertion level is popped off the stack. This is
consistent with the behavior specified by SMT-LIB2 when
\lstinline|:global-declarations| is \lstinline|false|. To be clear,
attempting to declare a variable with the same name and a different
sort as one in the current assertion level or any assertion level
below it will result in an error. The one exception is the
introduction of bound variables, which behave in the same way that
SMT-LIB2 describes above (any bindings with the same name as a bound
variable are replaced in the context of the body of the form
introducing the bound variables).

Our wrapper around the Z3 solver object contains an \emph{environment
  stack} that maps identifiers to variable declarations at each
assertion level. \lispsmt\ provides two ways to introduce variable
declarations: an ahead-of-time option (consistent with SMT-LIB2) and
an inline option. The ahead-of-time option involves using the
\lstinline|declare-const| or \lstinline|declare-fun| forms, which
behave identically to the commands of the same name defined by
SMT-LIB2. After checking that the variable is not already declared in
the current assertion level or any level below it, a variable declared
using either form is added to the solver's environment stack at the
current assertion level. The variable can then be referred to in any
assertions added at the current assertion level or above it. The
inline option for declaring variables involves providing variable
specifiers in a \lstinline|z3-assert| form. These variable specifiers
are processed to produce a mapping from each variable to a
declaration, and the declarations are added to the solver's
environment stack at the current assertion level. Just as with the
ahead-of-time option, an attempt to declare a variable with a name
that is already mapped to a declaration but a different sort than that
declaration will result in an error.

\paragraph*{Using and Defining Sorts}

In SMT-LIB2, each sort is defined with some number of parameters
(potentially zero). For example, the \lstinline|Int| sort takes in
zero parameters, and the \lstinline|Seq| sort takes in a single
parameter (representing the sort of the sequence's elements). A sort
with at least one parameter is a \emph{parametric} sort, while a sort
with no parameters is a \emph{non-parametric} sort. The name of a sort
may be an \emph{indexed identifier}, meaning that it is of the form
\lstinline|(_ <name> <idx1> ... <idxn>)|, where each
\lstinline|<idx|$i$\lstinline|>| is either a number or a symbol. For
example, a sort representing a bitvector of width 3 is represented
using the following indexed identifier: \lstinline|(_ BitVec 3)|. This
is a non-parametric sort, though it may look like a parametric
one.


So, why does this distinction exist between indexed identifiers and
parametric ones? In SMT-LIB2, it is possible to define a \emph{sort
  parameter}---a variable that ranges over sorts---and then to use
that variable as a parameter for a parametric
sort. Listing~\ref{lst:sort-parameter-ex} shows an example where
\lstinline|X| is declared as a sort parameter, and then a variable
\lstinline|y| is declared to be a function from \lstinline|Int| values
to values of sort \lstinline|X|.
%
%
On the other hand, indexed
identifiers are restricted so that the provided indices are literal
values. This means that for any bitvector sort in a SMT-LIB2 query,
the width of that sort is encoded syntactically, making it much easier
to apply any analysis that might benefit from knowledge of the
bitwidth.

\begin{lstlisting}[style=acl2, label=lst:sort-parameter-ex, caption={An example highlighting how sort parameters can be expressed in SMT-LIB2 syntax.}]
;; X is a variable over sorts
(declare-sort-parameter X)
;; y is a variable over functions from Int to X
(declare-const y (-> Int X))
\end{lstlisting}

When declaring variables for use in assertions, it is necessary to
provide sort specifiers to indicate what sort each variable should
have. A sort specifier refers to either a non-parametric sort, a
parametric sort, or a function rank. We will first discuss
non-parametric and parametric sorts before discussing function ranks.

A sort specifier for a non-parametric sort is simply a symbol denoting
the name of a non-parametric sort that is known to \lispsmt. The
package of that symbol does not matter. For example, the SMT-LIB2 sort
\lstinline|Int| is known to \lispsmt, and both \lstinline|:int| and
\lstinline|int| are sort specifiers denoting it. A sort specifier for
a parametric sort is a list where the first element is a symbol
indicating the name of the parametric sort and the remaining entries
in the list are arguments for the parameters of the parametric
sort. Different parametric sorts may take in different kinds of values
for their parameters, including sorts. Parameters for such sorts can
be provided as sort specifiers themselves. For example,
\lstinline|(:seq :int)| is a sort specifier that denotes the
\lstinline|(Seq Int)| sort in Z3.

SMT-LIB2 requires that each function have at least one \emph{rank}
associated with it~\cite{smtlib}. A rank is a non-empty sequence of
sorts, where the last sort is the return type of the function and the
sequence of sorts up to the last sort (potentially empty) denotes the
sorts of the parameters of the function. A sort specifier for a
function rank consists of a list of the form
\lstinline|(:fn (<p1> ... <pn>) <r>)|, where each
\lstinline|<p|$i$\lstinline|>| and \lstinline|<r>| is a sort specifier
for a parametric or non-parametric sort. In general a function may
have multiple ranks, but \lispsmt\ only supports free functions with a
single rank, for similar reasons as it does not support variables that
have the same name but different sorts. Function rank sort specifiers
are processed by translating each of \lstinline|<p|$i$\lstinline|>|
and \lstinline|<r>| into Z3 sorts and then producing a Z3 function
declaration object with the given name and rank.

When declaring variables inline using \lstinline|z3-assert|, it is
necessary to refer to the name of the variable's sort using a keyword
symbol (which can be written as a symbol whose name starts with a
colon). This is because the fact that a symbol is in the keyword
package is used to identify that a particular entry in the variable
specifiers for a \lstinline|z3-assert| call refers to a type rather
than a variable name. When using \lstinline|declare-const| or
\lstinline|declare-fun|, the name of the sort will be normalized in
such a way that the package that it is in is irrelevant.

Many sorts built-in to Z3, like \lstinline|Int| and \lstinline|Seq|,
are available with the same names in \lispsmt. In addition, it is
possible to define a subset of the user-defined sorts that Z3
allows. In particular, \lispsmt\ supports enumeration sorts and tuple
sorts. Enumeration sorts consist of a finite number of distinct
constants. For example, one way to represent the value of a Sudoku
square is as an enumeration sort containing only the integers between
1 and 9 inclusive. Such an enumeration sort can be defined in
\lispsmt\ using the \lstinline|register-enum-sort| function, as shown
in Listing~\ref{lst:sudoku-enum-sort}.

\begin{lstlisting}[style=acl2, label={lst:sudoku-enum-sort}, caption={An example of an enumeration sort being registered in \lispsmt.}]
(register-enum-sort :square (1 2 3 4 5 6 7 8 9))
\end{lstlisting}

Tuple sorts can be thought of like \lstinline|struct| types in Common
Lisp or record types in other languages. They consist of a set of
fields, each of which has a name and a sort. The fields must have
distinct names. An example of a definition of a tuple sort is shown in
Listing~\ref{lst:example-tuple-sort}.

\begin{lstlisting}[style=acl2, label={lst:example-tuple-sort}, caption={An example of a tuple sort being registered in \lispsmt.}]
(register-tuple-sort :person ((age . :int) (name . :string)))
\end{lstlisting}

Both enumeration sorts and tuple sorts can be defined using the
\lstinline|declare-datatypes| SMT-LIB2 command, though that command
allows for the definition of more complicated sorts than \lispsmt\
does.

\paragraph*{Interpreting Models}

When Z3 determines that the set of assertions loaded into the current
solver is satisfiable, it is possible to request a \emph{model} from
Z3 that describes a satisfying assignment to the set of
assertions. This model maps any free variables and sorts in the
assertions to interpretations (values). However, Z3 may determine that
the interpretation of a particular free variable does not impact the
satisfiability of the assertions. The generated model will not provide
an interpretation for such variables.

\begin{sloppypar}
To be able to use the interpretations from a model in Common Lisp, it
is necessary to translate them into Common Lisp values. This can be
done by using the \lstinline|(get-model-as-assignment)| function. The
interpretations for constants are encoded as Z3 AST values, just like
the AST values that we generate when producing constraints for
asserting in Z3. For example, Z3 may encode an interpretation
equivalent to the sequence \lstinline|(1 2 3)| as a concatenation of
unit sequences
\lstinline|(seq.++ (seq.unit 1) (seq.unit 2) (seq.unit 3))|.  The code
that translates these ASTs into Common Lisp values does not support
all possible interpretations. Additionally, there are some ASTs that
have multiple possible Common Lisp representations, or for which it is not
possible to produce a perfectly identical Common Lisp representation
using Z3's C API. These include algebraic number values
representing irrational roots of polynomials (for example,
$\sqrt{2}$)---\lispsmt\ will by default translate such values into a
Common Lisp floating-point value, losing some precision.
\end{sloppypar}

Function interpretations are particularly interesting to look at. Z3
represents an interpretation for a function $f$ using a combination of
a map from function inputs to outputs $M_f$ and a default value
$\mathit{else}_f$. The value of the function on a particular set of
arguments $f(a_1, ..., a_n)$ is either the value that the set of
arguments maps to in $M_f$ (if that set of arguments is mapped by
$M_f$) or the default value $\mathit{else}_f$ otherwise. By default,
\lispsmt\ will translate a function interpretation for a function $f$
into an S-expression containing a map with all of the entries from
$M_f$ but where each argument value and output value has been
transformed into a Common Lisp value, plus a designated
\lstinline|:default| key that is mapped to $\mathit{else}_f$
transformed into a Common Lisp value.

As a result of the restrictions that \lispsmt\ imposes on variable
declarations, it is always the case that a model produced by Z3 will
contain at most one interpretation for each variable name. This makes
it possible to unambiguously interpret the assignment produced by
\lstinline|(get-model-as-assignment)|, as otherwise it would be
possible for multiple variables for the same name but different sorts
to be included in the assignment. On the other hand, it is possible
that some of the variables that were constrained will not appear in
the assignment produced by \lstinline|(get-model-as-assignment)|. This
occurs when Z3 does not assign the variable an interpretation in the
produced model.

Another way to interact with the model produced by Z3 is to use the
\lstinline|eval-under-model| form provided by \lispsmt. This form
takes in a expression to be converted into a Z3 AST in the same way
that \lstinline|z3-assert| does and evaluates it under either the
given model or the result of \lstinline|(get-model)| if no model is
provided. The result of the evaluation is another Z3 AST, which is
converted into a Common Lisp value and returned. By default
\lstinline|eval-under-model| will ask Z3 to perform completion on the
given model when evaluating the given expression, meaning that if the
statement to evaluate references a variable that was used in the
assertion stack but that is not assigned an interpretation in the
given model, Z3 will assign that variable some value that satisfies
its sort (and it will use this value consistently if the variable
appears more than once in the statement to evaluate).

\paragraph*{Additional Features}

Z3 provides a wide variety of features, many of which have not been
discussed so far. \lispsmt\ supports Z3's optimization functionality,
allowing one to specify objective functions to maximize or minimize as
well as add soft constraints. It is possible to access statistics that
Z3 gathers during the solving process, something which is often
helpful when trying to understand Z3's performance on a particular set
of assertions.

\section{Sudoku}
\label{sec:sudoku}

\begin{figure}
  \centering
  \begin{minipage}{0.45\textwidth}
    \includegraphics[width=\textwidth]{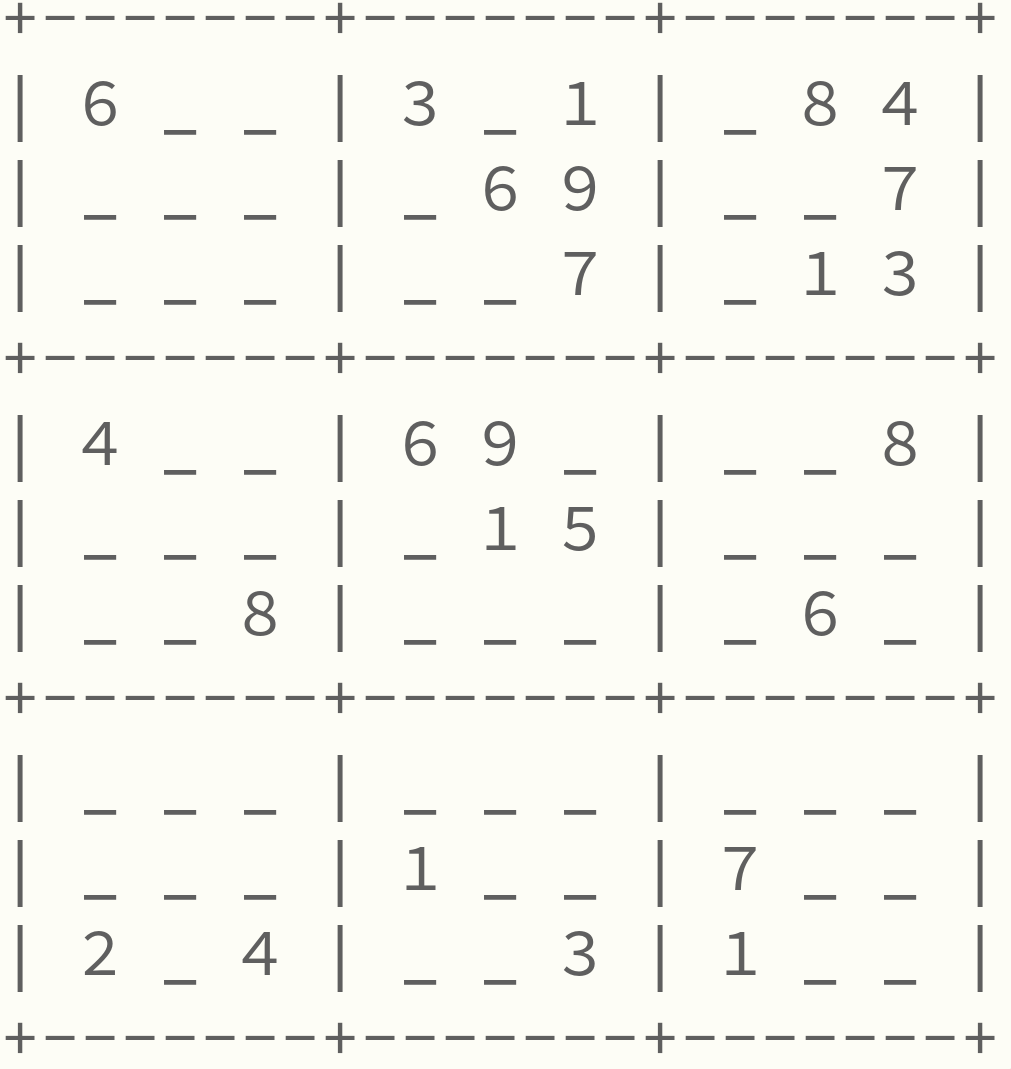}
  \end{minipage}
  \qquad
  \begin{minipage}{0.45\textwidth}
    \includegraphics[width=\textwidth]{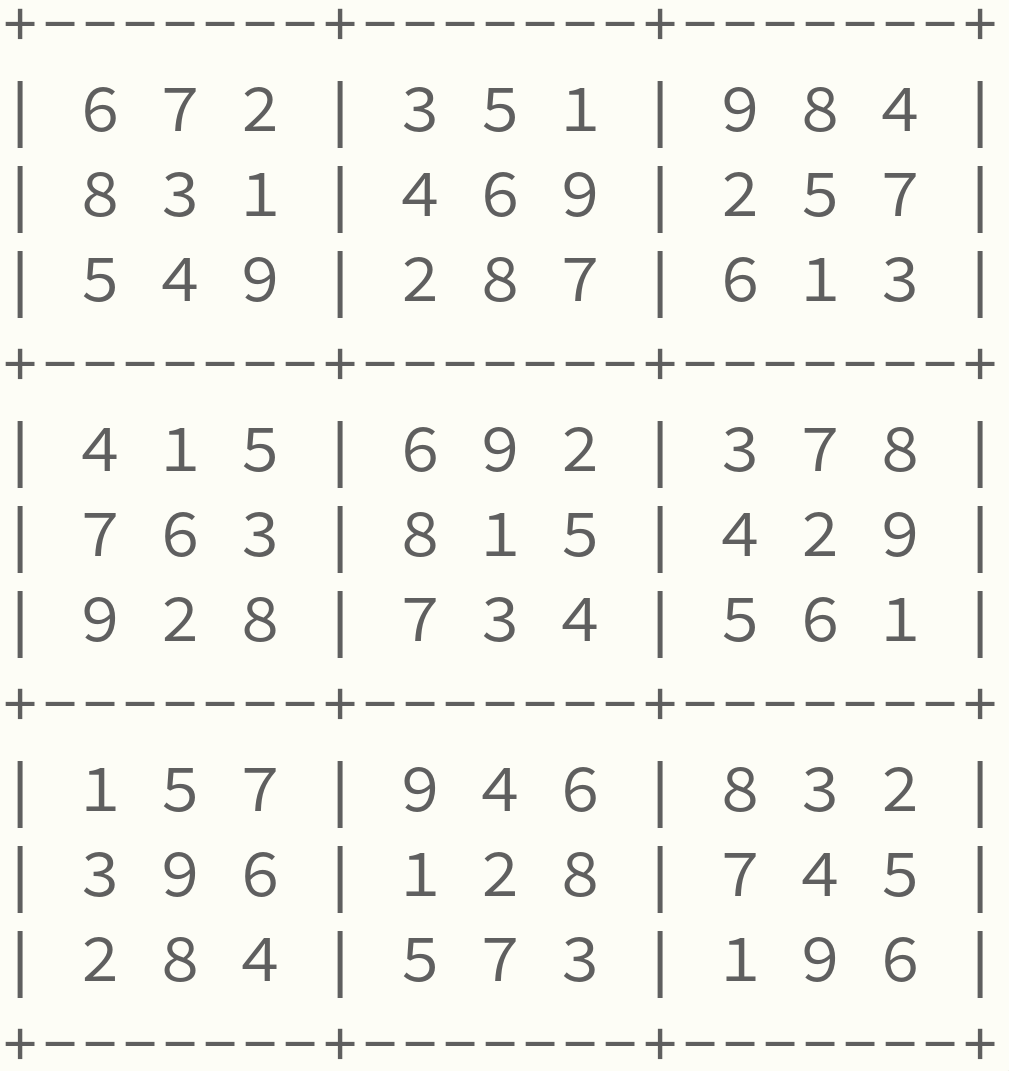}
  \end{minipage}
  \caption{A Sudoku puzzle (left) and its solution (right). _ is used to denote a blank square.}
  \label{fig:sudoku}
\end{figure}


A classic example of a puzzle that can be encoded as an SMT problem is
Sudoku. A traditional Sudoku puzzle consists of a 9x9 grid of squares
(a \emph{grid}), where each square is either blank or contains an
integer between 1 and 9 inclusive. The grid is partitioned into nine
3x3 submatrices (\emph{subgrids}). The goal of the player is to fill
in any blank squares such that the resulting grid satisfies the
following: for each row, column and subgrid, that group of squares
must contain distinct values. A well-formed Sudoku puzzle has a unique
solution~\cite{science-behind-sudoku}. Figure~\ref{fig:sudoku} shows
an example of a Sudoku puzzle and its solution.

One can encode Sudoku as an SMT problem using a variety of different
representations, but here we show one: representing the value of each
square using an integer variable, appropriately constrained to be
between 1 and 9 inclusive. Listing~\ref{lst:sudoku} contains an
implementation of a Sudoku solver using \lispsmt. It consists of
Common Lisp functions that generate the needed constraints for each
square, row, column and subgrid, which can be asserted in Z3 once when
the program starts up. There is also a function that translates a
representation of a Sudoku problem (a ``starting grid'') into a set of
equality constraints that represent the values of non-blank squares
given in the problem, which can then be asserted in Z3. This example
highlights Z3's scope functionality by first asserting the constraints
that are constant across all Sudoku problems (the square, row, column
and subgrid constraints), and then creating a new scope before adding
the constraints for a particular Sudoku problem. This new scope can
then be exited after the Sudoku problem is solved so that the Z3
solver can be used again for another Sudoku problem without needing to
re-assert the initial set of constraints.

\begin{lstlisting}[style=acl2, label=lst:sudoku, caption={An implementation of a Sudoku solver using \lispsmt. This is an excerpt from the Sudoku example that is provided with \lispsmt, which also includes pretty-printing and several Sudoku puzzles.}]
;; Turn an index into a Sudoku grid into the variable corresponding to that square's value.
(defun idx-to-cell-var (idx)
  (intern (concatenate 'string "C" (write-to-string idx))))

;; We'll encode the sudoku grid in the simplest way possible, 81 integers
(defconstant +cell-vars+
  (loop for idx below 81
        append (list (idx-to-cell-var idx) :int)))

;; We limit the integers to values between 1 and 9, inclusive
(defconstant cell-range-constraints
  (loop for idx below 81
        append `((<= 1 ,(idx-to-cell-var idx))
                 (>= 9 ,(idx-to-cell-var idx)))))

;; distinct is a built-in Z3 function that is true iff none of its arguments are equal.

;; The values in each row must be distinct
(defconstant row-distinct-constraints
  (loop for row below 9
        collect `(distinct
                  ,@(loop for col below 9
                          collect (idx-to-cell-var (+ (* 9 row) col))))))

;; The values in each column must be distinct
(defconstant col-distinct-constraints
  (loop for col below 9
        collect `(distinct
                  ,@(loop for row below 9
                          collect (idx-to-cell-var (+ (* 9 row) col))))))

;; The values in each 3x3 box must be distinct
(defconstant box-distinct-constraints
  ;; These numbers are the indices of the top-left square of each box
  (loop for box-start in '(0 3 6 27 30 33 54 57 60)
        collect `(distinct
                  ;; These numbers are the offsets of each square in a
                  ;; box from the index of the box's top-left square
                  ,@(loop for box-offset in '(0 1 2 9 10 11 18 19 20)
                          collect (idx-to-cell-var (+ box-start box-offset))))))

;; Set up the initial constraints on the grid
(defun init ()
  (solver-init)
  ;; z3-assert-fn allows us to assert an expression generated by executing some Common Lisp code.
  (z3-assert-fn +cell-vars+ (cons 'and cell-range-constraints))
  (z3-assert-fn +cell-vars+ (cons 'and row-distinct-constraints))
  (z3-assert-fn +cell-vars+ (cons 'and col-distinct-constraints))
  (z3-assert-fn +cell-vars+ (cons 'and box-distinct-constraints)))

;; This generates constraints based on a "starting grid".
;; This starting grid is a length-81 list representation of the 9x9 Sudoku grid in row-major order.
;; The list should have a _ in cells where no initial value is given.
(defun input-grid-constraints (grid)
  (loop for entry in grid
        for idx below 81
        when (not (equal entry '_))
        collect `(= ,(idx-to-cell-var idx) ,entry)))

(defun solve-grid (input-grid)
  (solver-push)
  (z3-assert-fn +cell-vars+ (cons 'and (input-grid-constraints input-grid)))
  (let* ((sat-res (check-sat))
         (res (if (equal sat-res :sat)
                  (get-model-as-assignment)
                sat-res)))
    (progn (solver-pop)
           res)))

;; Now, use the solver! We assume the Sudoku grid from Figure 1 is loaded in *fig1-grid*.
(init)
(solve-grid *fig1-grid*)
\end{lstlisting}


Generating constraints programatically makes it easy to experiment
with variants of Sudoku that have different-sized grids. The
traditional Sudoku game seen above can be thought of as 3x3
Sudoku---each subgrid is contains 3 rows and 3 columns of squares and
the top-level grid contains 3 rows and 3 columns of subgrids. The
\lispsmt\ examples contain code for a solver that can solve $n$x$n$
Sudoku problems.

Note that the above Sudoku solver does not make use of any \acls\
functionality. However, one could imagine using this Sudoku solver
implementation as an oracle for solutions to a Sudoku problem inside
of \acls. Its output need not be trusted---instead one could write
\acls\ functions to validate that the produced solution is indeed a
valid solution and matches with the given Sudoku problem.

\section{Application: String Solving}
\label{sec:string-solving}




Several applications benefit from the ability to perform
satisfiability checking over string equations (``string
solving''). These include security
analysis~\cite{string-constraint-solver-webapp, hampi-string-solver},
program verification~\cite{evaluation-string-solvers,
  string-analysis-book} and type
checking~\cite{remora-constraint}. Many string solvers exist,
including Z3str3~\cite{z3str3} which is part of Z3.


Kumar and Manolios used \lispsmt\ in their string solver
SeqSolve~\cite{ankit-mpmt}. In addition to supporting string equation
constraints, SeqSolve allows one to express LIA constraints over the
lengths of string variables. As part of its solving process, SeqSolve
generates LIA constraints over the lengths of string variables and
uses \lispsmt\ to determine whether or not these constraints are
satisfiable.
SeqSolve uses Z3's incremental solving capabilities (\eg\ the
assertion stack) to manage adding and removing constraints as
appropriate as the string solving algorithm partitions the search
space and explores each partition. SeqSolve is a particularly good
example of the advantages of ASPF, as it is partially written in
ACL2s, enabling it to take advantage of defdata data definitions and
to express guarantees regarding those functions written in ACL2s. This
includes guarantees about types (the function always returns a value
satisfying its signature if it was called with arguments satisfying
its signature) as well as regarding termination (unless explicitly
configured not to, \acls\ requires that any admitted function
terminates).

Kumar and Manolios evaluated SeqSolve on a set of benchmarks,
comparing against a set of string solvers that also supported length
constraints. The benchmark set was derived mainly from benchmarks used
in prior work, omitting benchmarks outside of the theory that SeqSolve
supports. Kumar and Manolios found that SeqSolve solved a larger
number of the benchmark problems in a shorter time than any of the
other string solvers at the time of their paper's publishing. The
results of Kumar and Manolios' work~\cite{ankit-mpmt} highlight the
ability of \lispsmt\ to enable the use of Z3 alongside \acls\ in a
performant way, and their feedback was invaluable in guiding continued
improvements to \lispsmt.

\section{Application: Wi-Fi Fuzzing}
\label{sec:wifi}

Wireless communication protocols are ubiquitous in modern life,
connecting devices ranging from smartphones to medical implants to the
Internet. One of the most prevalent wireless communication protocols
is the IEEE 802.11 Wi-Fi protocol~\cite{802-11-spec}. Given the reach
and impact of devices implementing Wi-Fi and the complexity of the
protocol, it is important to evaluate the correctness and security of
Wi-Fi infrastructure that will be used in sensitive applications. It
is for this reason that we collaborated with a group at Collins to
work on hardware-in-the-loop fuzzing of Wi-Fi
routers~\cite{enumerative-data-types}.

Fuzzing is a technique for software testing wherein the system under
test (SUT) is run on generated inputs with the goal of evaluating its
reliability. Hardware-in-the-loop fuzzing involves using fuzzing to
test a physical hardware device. This introduces several challenges
above and beyond software fuzzing, including the introduction of
timing constraints. These challenges, in conjunction with the
complexity of the Wi-Fi protocol itself, mean that fuzzing Wi-Fi
devices is challenging. Our work focused primarily on a particular
part of the 802.11 specification---the probe request frame, which is
used to request information about a Wi-Fi router prior to connecting
to it. A 802.11 frame contains several fields, one of which is a body
that itself consists of a sequence of elements. These elements each
contain an ID indicating the kind of the element (are chosen from a
set of possible element kinds) and a body, the valid values of which
are determined by the element's kind.

Our collaborators had been working on model-based fuzzing of Wi-Fi
routers, focusing on the probe request frame. They wrote a model
expressing the structure of and constraints on a probe request frame
and (through some tooling) translated the model into a SMT query, a
satisfying assignment for which corresponded to a valid probe request
frame. They then used an approach called trapezoidal
generalization~\cite{trapezoidal-generalization} to generate a large
number of different probe request frames from a single satisfying
assignment. This was necessary because solving the SMT query was quite
slow. The probe request frames were then sent to the SUT, which was
monitored for availability.

We were interested in comparing our collaborators' approach for
model-based fuzzing against one using ACL2s' enumeration
facilities. To do this, we translated our collaborators' model into
data definitions in ACL2s using the defdata system. By doing so, we
were able to make use of the \emph{enumerators} that ACL2s generates
for each defdata type. The enumerator for a type is a function that
maps natural numbers to elements of the type, making it possible to
generate arbitrary elements of the type. This is used inside of ACL2s
to support counterexample generation and automated testing.

An important factor for the success of model-based fuzzing is the
ability to explore a large swathe of the space of possible input for
the SUT (in the case of a Wi-Fi router, the space of possible
frames). One relevant parameter in the case of probe request frames is
the size of the generated frames, computed by summing the size of each
element in the frame's body with the size of the frame's header. There
exist valid probe request frames with any size between 172 and 2741
bytes (inclusive). It is possible that a bug in a Wi-Fi router's
handling of probe request frames may only occur given very large or
small frames. Therefore, we evaluated the ability of ACL2s to generate
probe request frames with a particular size against an SMT
approach. We found that both struggled to generate probe request
frames across the entire range of valid sizes. We determined that the
reason why ACL2s struggled to generate frames with a particular size
was because it struggled to reason about the size of the different
parts of a frame independently from their contents and pass that
information to the frame's enumerator.

To produce a more performant approach, we split the task of generating
a frame with a particular size into two steps: we solve the size
constraints first (in Z3) and then pass that information along to the
appropriate enumerators in ACL2s. This leverages the strengths of each
tool---Z3 is highly capable at solving constraints involving linear
integer arithmetic, and ACL2s makes it easy to specify types
describing the body of each element, as the constraints on the
contents of the body can be difficult for Z3 to handle.

\begin{figure}[t]
  \centering
\includegraphics[width=0.8\textwidth]{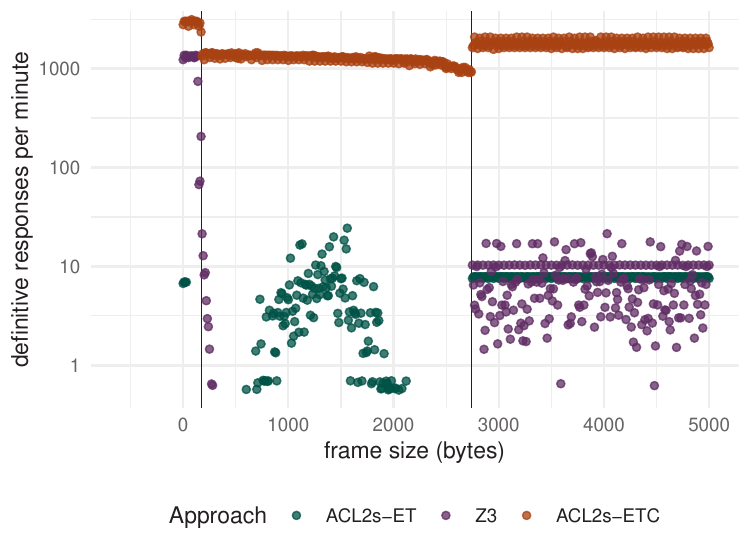}
\caption{The number of frames generated per minute using each of three
  approaches when queried for frames with a given length. Z3 denotes
  an implementation that uses Z3 and an SMT-LIB2 version of the model
  to generate frames, ACL2s-ET uses ACL2s' counterexample generation
  and ACL2s-ETC uses a combination of Z3 and ACL2s. Only instances
  where the model returned a definitive response (\eg\ not ``unknown''
  or ``timeout'') are shown. The two vertical lines represent the
  minimum frame size and the maximum frame size; any responses outside
  of that range were all UNSAT, and any within that range were
  SAT. This figure is similar to Figure 7 in
  \cite{enumerative-data-types}.}
\label{fig:fpm-plot}
\end{figure}

To evaluate the performance of our approach (which we will refer to as
ACL2s-ETC), we compared it against an approach that purely used ACL2s'
counterexample generation (referred to as ACL2s-ET) and one that
purely used Z3. We already had the ACL2s version of the model and we
translated the model into SMT-LIB2 constraints for use in evaluating
the pure-Z3 approach. We then measured the throughput of the three
approaches when queried for frames with a particular length. We
evaluated across lengths from 0 to 5000 bytes, in increments of 10
bytes. Given the model, frames exist with sizes between 172 and 2741
bytes inclusive, so the evaluated range contained both satisfiable and
unsatisfiable queries. The results are shown in
Figure~\ref{fig:fpm-plot} (note that results for frame sizes between
4000 and 5000 bytes are elided).

The results show a clear win for ACL2s-ETC: it was able to generate
frames more quickly than either ACL2s-ET or Z3 across the entire range
of evaluated frame sizes, and much more quickly (nearly two orders of
magnitude) across the range of satisfiable frame sizes. These results
are strong evidence for the effectiveness of the combination of ACL2s
and Z3 and of the ability of \lispsmt\ to support such a
combination. Further discussion of these results is available in our
FMCAD paper~\cite{enumerative-data-types}.



\subsection{Enumerative Data Types Modulo Theories}

Our evaluation above showed the effectiveness of our ACL2s-ETC
approach in a particular application, but implementing our approach
there required a fair amount of manual effort with respect to plumbing
together defdata enumerators and Z3. We would like to generalize our
approach in such a way that users can take advantage of it in a highly
automated fashion. This is the main idea behind our ongoing work on
\emph{enumerative data types modulo theories} (EDT), an extension of
defdata that allows one to express what we call \emph{parameters} of
data types---features that can be constrained and solved for using
something like Z3. For example, in our Wi-Fi fuzzing work we would
define a frame size parameter which could then be solved for ahead of
time if the user wrote a constraint over its value. We hope to use
\lispsmt\ to power our implementation of EDT inside of ACL2s' cgen
counterexample generation functionality.

\section{Related Work}
\label{sec:related-work}

The ACL2 Sedan (ACL2s)~\cite{dillinger-acl2-sedan, acl2s11} is an
extension of the ACL2 theorem prover\cite{acl2-car, acl2-acs,
acl2-web}. On top of the capabilities of ACL2, ACL2s provides the
following:
(1) A powerful type system via the defdata data definition
framework~\cite{defdata} and the \texttt{definec} and
\texttt{property} forms, which support typed definitions and
properties.
(2) Counterexample generation capability via the cgen
framework, which is based on the synergistic integration of theorem
proving, type reasoning and testing~\cite{cgen, harsh-fmcad,
  harsh-dissertation}.
(3) A powerful termination analysis based on calling-context
graphs~\cite{ccg} and ordinals~\cite{ManoliosVroon03, ManoliosVroon04,
  MV05}.
(4) An (optional) Eclipse IDE plugin~\cite{acl2s11}.
(5) The ACL2s systems programming framework
(ASPF)~\cite{acl2s-systems-programming} which enables the development
of tools in Common Lisp that use ACL2, ACL2s and Z3 as a
service~\cite{enumerative-data-types,
  invariant-discovery-game, ankit-mpmt, acl2-workshop-checker-paper}.


The most directly relevant existing system to \lispsmt\ is the
Smtlink tool, developed by Peng and Greenstreet~\cite{smtlink,
  smtlink-two}. In short, Smtlink allows one to use SMT to discharge
ACL2 goals in a way that only involves trusting a small amount of code
(in addition to ACL2 and Z3). In more detail, Smtlink provides a set
of verified clause processors for transforming ACL2 goals into
equivalent forms that are better suited for the SMT solver and a
trusted clause processor that translates an ACL2 goal into a set of
constraints for Z3, calls Z3, and interprets the output. Smtlink
provides support for reporting counterexamples if reported by Z3 as a
result of a failed proof.

Part of the challenge of implementing a system like Smtlink is that of
providing a translation of an ACL2 form into Z3 that accounts for the
differing semantics of ACL2 and Z3. This is especially true given that
ACL2 uses an untyped logic and Z3 uses a many-sorted logic. For
example, consider the ACL2 theorem in
Listing~\ref{lst:acl2-fixing-example} which can be proven to hold.

\setcounter{footnote}{-1}

\begin{lstlisting}[style=acl2, label=lst:acl2-fixing-example, caption={A theorem
    expressed using ACL2s' property form. Note that for ACL2s
      to accept this property, the contract checking feature of
      \lstinline|property| must be disabled.
      \protect\footnote{Contract checking can be disabled by evaluating \lstinline|(set-acl2s-property-table-test-contracts? nil)| and \lstinline|(set-acl2s-property-table-check-contracts? nil)|}
    }]
(property (x :bool y :int)
  (= (+ x y) y))
\end{lstlisting}

This is true in ACL2 since ACL2's arithmetic functions generally treat
any non-number arguments as though they were 0. However, the
corresponding Z3 expression is falsifiable! This is shown in
Listing~\ref{lst:z3-fixing-example}. The reason for this is that Z3
happens to treat \lstinline|true| as 1 and \lstinline|false| as 0 in
the context of arithmetic operators \footnote{Note that this behavior
  is an extension of the behavior that SMT-LIB requires, and other SMT
  solvers supporting SMT-LIB may handle things differently. For
  example, CVC5 reports an error when trying to add that assertion, as
  it does not define the addition and multiplication operators on
  Booleans.}.

\begin{lstlisting}[style=acl2, label=lst:z3-fixing-example, caption={A na{\"i}ve translation of the theorem in Listing~\ref{lst:acl2-fixing-example} into \lispsmt\ calls.}]
;; We negate the statement that we are trying to prove, and if
;; Z3 determines UNSAT then the statement is valid.  
(z3-assert (x :bool y :int)
  (not (= (+ x y) y)))
(check-sat) ;; returns :sat, therefore the statement is not valid.
\end{lstlisting}

Since Smtlink allows one to use the fact that Z3 can prove a
transformed ACL2 proof obligation to justify the correctness of the
obligation in ACL2, a difference in semantics between the two that
is not accounted for can load to unsoundness. The authors of Smtlink
took particular care to ensure that their translation between ACL2 and
Z3 expressions preserve validity. We chose to develop a lighter-weight
solution in \lispsmt, insofar as it does not provide the ability to
translate of an ACL2 expression into a Z3 expression in a way that
preserves validity.

Manolios and Srinivasan were early advocates for integrating decision
procedures into interactive theorem provers, suggesting an integration
of the UCLID decision procedure with ACL2 in
2004~\cite{04-manolios-hard}. This integration was later performed and
used to verify pipelined processor models~\cite{verifying-bit-level,
  framework-verifying-pipelined-machines}, enabling the automated
verification of proofs that neither UCLID nor ACL2 could handle
alone. Srinivasan went on to develop an integration of the Yices SMT
solver with ACL2 as part of his PhD
thesis~\cite{sudarshan-thesis}. Other researchers have investigated
the integration of SMT into the
Isabelle/HOL~\cite{expressiveness-automation-soundness} and
Coq~\cite{coq-smt-integration} theorem provers, and work is ongoing to
integrate the CVC5 SMT solver~\cite{cvc5} and the Lean theorem
prover~\cite{challenges-smt-proof-production}.

\section{Conclusion and Future Work}
\label{sec:conclusion}

We presented \lispsmt, an extension to the ACL2s systems programming
framework (ASPF) that supports the use of Z3 as a service. The source
code for \lispsmt\ plus documentation and several examples of its
usage are publicly available~\cite{repo}. We also discussed
three applications of our extended ASPF, the first being a Sudoku
solver and the second being the SeqSolve string solver. The last
application involved testing of wireless routers, where using a
combination of ACL2s and Z3 resulted in substantially improved
performance over pure-ACL2s and pure-Z3 approaches. We expect to use
\lispsmt\ inside of ACL2s as part of our ongoing work on enumerative
data types modulo theories.

There are many improvements that we would like to make to
\lispsmt. These include supporting a larger subset of the commands,
operators and sorts that Z3 and SMT-LIB2 provide, developing an
optional integration between Z3 sorts and ACL2s defdata types and
enabling the use of other SMT solvers on the backend (in particular,
CVC5). We are interested in getting more feedback from external users
of the interface and encourage anyone interested in using \lispsmt\ to
experiment with it and reach out with any questions, comments or
feedback.

\paragraph*{Acknowledgments}
\lispsmt\ has been greatly improved thanks to feedback from its users,
including Ankit Kumar and the students in the Fall 2021 and 2022
sections of CS4820 at Northeastern University. Additionally, we would
like to thank David Greve, who collaborated on Wi-Fi fuzzing with us
at Collins, as well as Konrad Slind, Kristopher Cory and all of the
other folks at Collins who we worked with.

\bibliographystyle{eptcs}
\bibliography{refs}
\end{document}